\documentclass[aps,showpacs,twocolumn]{revtex4} 
\usepackage{graphicx,tikz}
\usepackage{amsmath}
\usepackage{tikz}
\usetikzlibrary{patterns}
\def \be{\begin{equation}}
\def\ee{\end{equation}}
\def\bea{\begin{eqnarray}}
\def\eea{\end{eqnarray}}
\usetikzlibrary{positioning}
\definecolor{bronze}{rgb}{0.8, 0.5, 0.2}
\definecolor{ao}{rgb}{0.0, 0.0, 1.0}
\def\be{\begin{equation}}
\def\ee{\end{equation}}
\def\bea{\begin{eqnarray}}
\def\eea{\end{eqnarray}}
\def\bd{\begin{displaymath}}
\def\ed{\end{displaymath}}

\def\etal{{\em et al. }}

\begin{document}
\title{Neutron capture reactions relevant to s-process and p-process in the domain of the $N=50$ shell closure}
\author{Saumi Dutta}
\email{saumidutta89@gmail.com}
\author{G. Gangopadhyay}
\email{ggphy@caluniv.ac.in}
\author{Abhijit Bhattacharyya}   
\email{abhattacharyyacu@gmail.com} 
\affiliation{Department of Physics, University of Calcutta\\
92, Acharya Prafulla Chandra Road, Kolkata-700 009, India}
\textwidth 7in
\textheight 8in

\begin{abstract}
Radiative thermal neutron capture cross sections for nuclei participating in s-process and p-process nucleosynthesis in and around $N=50$ closed neutron shell have been calculated in statistical semi-microscopic Hauser-Feshbach approach for the energy range of astrophysical interest. A folded optical-model potential is constructed utilizing the standard DDM3Y real nucleon-nucleon interaction. The folding of the interaction  with target radial matter densities,  obtained from the relativistic-mean-field approach, is done in coordinate space using the spherical approximation. The standard nuclear reaction code TALYS1.8 is used for  cross-section calculation.  The cross sections are compared with experimental results and reasonable agreements are found for almost all cases. Maxwellian-averaged cross sections (MACS) for the nuclei are presented at a single thermal energy of  30 keV relevant to s-process. We have also presented the MACS values  over a range of energy from 5 to 100 keV for neutron magic nuclei with $(N=50)$.
\end{abstract}
\maketitle

\section{introduction}
Elements heavier than iron are produced via two principle processes, namely, the slow neutron
 capture process (s-process) and the rapid  neutron capture process (r-process), differing in the respective neutron capture timescales with respect to the $\beta$-decay half-lives.  There is  a minor contribution 
from another process, namely, p-process, producing a subset of
 proton-rich isotopes. 
 The detailed study of the heavy element nucleosynthesis was done in the fundamental
 work of Burbidge \etal \cite {B2FH} and also of Cameron \cite{cameron}. 
 Recently K\"{a}ppeler  \etal \cite{kappeler1} presented a review
 of progress on the studies of s-process nucleosynthesis with advanced nuclear physics inputs, observational data, and stellar models.

While the majority of the theory of the s-process is well-developed, uncertainty still remains in constraining the neutron capture rates of nuclei involved in nucleosynthesis chain. The capture cross sections are highly scattered and uncertain in the energy range  appropriate for astrophysical applications. These uncertainties lead to significant errors in determining exact abundances of elements involved in different processes. Many works have been devoted to this respect in order to measure the thermal neutron capture cross sections relevant to s-process temperature. However, reactions on some important nuclei are still
not available due to their unavailability in the terrestrial laboratory. K\"{a}ppeler \etal \cite{kappeler2} showed the present status of the uncertainty of stellar $(n,\gamma)$ cross sections and commented that improvements are certainly necessary especially in the mass region below A=120 and above A=180.   Some reactions are of significant importance, basically those with closed neutron shells acting as bottlenecks to the s-process reaction flow. They have very low cross-sections and in some scenarios, the s-process reaction flow cannot overcome these bottlenecks. The neutron capture cross sections of the branch-point nuclei in the s-process path also have to be known with better accuracy. The branching in the s-process path appears when a competition between neutron capture and $\beta$-decay takes place due to the appearance of a long-lived nucleus with a larger probability of neutron capture than $\beta$-decay. Further propagation depends on the neutron density and temperature and hence branching analysis can be used as an efficient monitor of stellar neutron density and temperature. These branchings also have an overall influence in changing the nucleosynthesis path of s-process and hence on the final yields of the nuclei produced. Thus, more accurate 
 cross sections are nowadays also needed to resolve the discrepancies in overproduction problems of some elements. 

 The  s-process is  subdivided into weak, main, and strong components. 
Recently,  another component, called lighter element primary process (LEPP) is also proposed.
 The neutron exposure for weak s-process is too low to achieve flow equilibrium 
and hence a small uncertainty in capture cross section may lead to significant 
changes in abundance for a number of heavier elements. This so-called propagation problem requires
 new data for the weak component in between Fe to Sr.
 Our present study deals  with nuclei near $N=50$ shell closure where mainly the weak s-process component  
 creates elements between Fe to Sr-Y-Zr region.  In a previous work, we studied the thermal $(n,\gamma)$ cross sections near $N=82$ shell closure
in between isotopes of xenon to samarium \cite{n82} which are produced by the main component of s-process nucleosynthesis. The astrophysical origins of the weak and main components are completely different. The weak component  occurs during the convective He core and C-shell burning in massive stars driving the  material with masses  56 $\textless$ A $\textless$ 90, in contrast to the main component, occurring mostly in asymptotic giant branch (AGB) stars. However, there remains some confusion about the marginal border between these two components and most of the nuclei considered in the present study reside in the vicinity of this border region. They are considered to be produced in both main and weak components of stellar s-process nucleosynthesis. For example, 50\% of the solar abundances of the s-only pair $^{86,87}$Sr are formed in the main component and remaining 50\% are from the weak component of s-process \cite{kappeler2}. In the concerned mass region, there are six stable nuclei on the nucleosynthesis path with  magic neutron number $N=50$, namely, $^{86}$Kr, $^{87}$Rb, $^{88}$Sr, $^{89}$Y,
and $^{90}$Zr, and $^{92}$Mo.  

The present study also involves some nuclei produced in astrophysical p-process near $N=50$ closed shell ($^{84}$Sr, $^{92,94}$Mo, $^{96,98}$Ru).  The p-nuclei, those are 10 to 100 times less abundant than s- or r-nuclei in the same mass range  are produced in the high-temperature environment via the sequences of photo-disintegration and $\beta^{+}$ decays driving the path through the extreme proton-rich side.  However, current theoretical models are incapable of reproducing the p-nuclei abundances in good agreements (except in certain mass range) with observation  and this inadequacy is directly related to the absence of proper nuclear inputs with reduced uncertainties.  Thus, $(n,\gamma)$ reactions have significant impact in p-process abundance determination 
as they hinder the flux via competing with $(\gamma,n)$ reactions. Some studies \cite{rauscher, rapp} have revealed a high degree of sensitivity towards the $(n,\gamma)$ rate variations in p-process during the network calculations. For example, Rayet \etal  found that the 
suppression of $(n,\gamma)$ channel  leads to  change the p-nuclei overproduction  factors, in some cases by order   as large as 10 or 100 \cite{rayet}. Hence, accurate determination of cross sections is highly needed for them.
 Particularly the $(n,\gamma)$ cross sections on p-nuclei are difficult to measure as they are not found in  significant amounts for time-of-flight (TOF) or activation measurements.  Experimental data are very scarce in this respect and by far most of the rates are to be inferred from statistical model calculations. 

In the domain  of $N=50$ shell proximity, some nuclei exist which are predominantly
 produced via s-process only. These so called s-only isotopes ($^{86}$Sr, $^{88}$Sr, and $^{96}$Mo) are of special 
importance as they can provide necessary clues to the s-process branchings. Hence, accuracy is also required in
 the cross sections of these s-only isotopes.

Some other important aspects, for which more accurate and enhanced nuclear data are required, are those of unstable elements having isotopic anomalies. For example, in the region of $N=50$ shell closure, Kr and Zr isotopes show anomalies  in meteoritic isotopic distribution.  There remain discrepancies in between measured and model-based  isotopic ratios and a spread in the ratio with grain size of the material is a persistent problem.
The isotopes $^{93}$Zr and $^{99}$Tc are of particular interest as they are being used in the nuclear transmutation of long-lived fission products (LLFP). The capture
cross sections of the stable zirconium isotopes as well as $^{93}$Zr are very
important for the transmutation study on $^{93}$Zr.    Similar to $^{93}$Zr, the isotope $^{99}$Tc with half-life of 2.11$\times$10$^{5}$ years and a large cumulative yield in the thermal neutron fission of $^{235}$U, is also one of the most important LLFPs for transmutation technology to convert the radioactive wastes into stable or short-lived nuclei.  The unstable isotopes are not available for direct measurement. Even the  experimental data on stable isotopes are inadequate and hence, theoretical estimates are strongly required for them.

It is also possible to predict the values of stellar mean
 neutron exposure relevant to s-process from the measurement of $(n,\gamma)$ cross sections. For example, Bauer \etal \cite {sr86sr87} 
recommended a set of neutron density, temperature, 
and mean neutron exposure from the cross section measurements, both for main and weak components. There are a few other works which
 investigated the same quantities from radiative cross sections \cite{kappeler1}. 
   The  $(n,\gamma)$ cross sections can also be used in the study of nuclear cosmochronology to determine the age of the Galaxy \cite{chrono1, chrono2, chrono3}. 
Though the neutron capture cross sections, in general, have $1/v$ dependence, this can significantly
differ when the p-wave capture is superimposed on pure s-wave contribution, thus resulting in an increase
 in the cross-section values with incident neutron energy.
 
 In the reaction path, neutron capture reactions on some nuclei populate both ground and isomeric states of the residual nuclei. Thus, it is also possible to measure the isomeric ratios of those elements if both total cross section and partial capture cross sections to isomeric states are known. Moreover, certain branchings in the reaction path depend on the probability of populating the isomeric states \cite{isomer_ratio}. The population of long-lived isomeric states can be important in branching analysis if they do not achieve thermal equilibrium in the stellar environment of s-process \cite{popul_isomer}. One such branch point in this region exists at $^{85}$Kr.

Besides the strong branch point at $^{85}$Kr, there exist some weaker branch points.
Weak branchings can be analyzed only if accurate MACS values are known.
  
Measurement of neutron capture reactions requires an energy resolution comparable to the 
distance between the overlapping levels of compound nucleus in which the captures occur. Recently much progress has been made in 
the neutron induced cross-section measurement. Bao \etal \cite{bao} have  recommended a large set of MACS values
 for $(n,\gamma)$ reactions relevant to s-process. We have also compared our MACS results with the theoretical MOST calculations. MOST is a Hauser-Feshbach code \cite{most1,most2}  which derives all nuclear inputs from global microscopic models \cite{most_inputs,n82}. 
There are various experimental techniques for measuring $(n,\gamma)$ cross sections, each suffering from its own inherent deficiency.
The most widely used methods are the time-of-flight (TOF)  and activation techniques.
 TOF is particularly  suitable for measuring $(n,\gamma)$ cross sections over a broad 
energy range which in turn is required in obtaining MACS values. However, more accurate measurements are in demand for various reasons. 
Some recent techniques, such as Pulse Height Weighing Technique (PHWT), n\_TOF and 4$\pi  \gamma$ ray detectors can give results with sufficiently reduced background and  neutron sensitivity. Still experiments are far away from producing precise measurements over the entire mass range. For example, reactions having pronounced contribution from direct capture processes  are hard to be measured via TOF detectors. DC mechanism is important mainly near magic neutron numbers. For example, capture on $^{88}$Sr possesses significant DC contribution. 

The paper is organized as follows. In the next section, we have discussed  theoretical background of our calculation. Next, in section III, we have discussed the results obtained from our theoretical approach. First, the   $(n,\gamma)$ cross sections for nuclei near  $N=50$  closed shell, those take part in the nucleosynthesis chain, are calculated with our theoretical approach and plotted with available experimental data.  The  MACS values, at a single thermal energy of 30 keV, are tabulated with experimental values and MOST2005  predictions. Further, MACS values for the energy range from  5 to 100 keV  are also presented for neutron magic nuclei.  Lastly, the summary is given followed by acknowledgment and bibliography.

\section{Theory}
\subsection{Relativistic-Mean-Field Model}
Relativistic-mean-field model is highly successful in describing binding energies, charge radii, different ground and excited state properties of nuclei, deformation, etc. We have chosen FSU Gold parameterization \cite{fsugold} for Lagrangian density \cite{lagrangian}.  We have obtained 
binding energies and charge radii in spherical RMF approach and compare with available experimental data. 
The difference between binding energies from theoretical RMF approach and experimental data $(\Delta_{\nu \pi})$ is attributed to the strength of n-p interaction of nucleus 
and also to some extent to the odd-even mass difference and a correction is done in this respect. This is  taken care of by the Casten factor P \cite{casten} defined as,
\begin{equation}
P=\frac {N_{p}N_{n}}{N_{p}+N_{n}}
\end{equation}
Where $N_{p}$ and $N_{n}$ are the number of valence particles (or holes past mid-shell) and the difference is given by 
aP, where a=-2.07 MeV \cite{castencorr1}.
More information can be available in Ref. \cite{castencorr1}. This correction 
has been carried out in the present work.

The point proton  density ($\rho_{p} $), obtained from relativistic-mean-field theory is folded with standard Gaussian form factor F({\bf r}) to obtain the charge density ($\rho_{ch}({\bf r})$).
\be
\rho_{ch}({\bf r})=\int \rho_{p}({\bf r\prime})F({\bf r}-{\bf r\prime}) d{\bf r\prime}
\ee
\be
F(r)=(a \sqrt\pi)^{-3}exp(-r^{2}/a^{2})
\ee
where, the constant $a=\sqrt{2/3}a_p$ with $ a_{p}$=0.8 fm, being the r.m.s radius of proton.
 The integration is done in coordinate space.

 The charge-density distribution thus obtained are compared with densities obtained from Fourier-Bessel (FB) parameter fit.
 FB parameters are derived from the experimental elastic electron scattering data \cite{devries}.
The charge densities are then determined using the following relation.
\[
    \rho(r)= 
\begin{cases}
    \sum_{n} a_{n} j_{0}(\frac{n\pi r}{R}),& \text{for r}\ll \text{R}\\
    0,              & \text{for r} \gg \text {R}
\end{cases}
\]
Here $a_{n}$ denotes the FB coefficients and $j_{0}(qr)$ denotes the Bessel function of order zero.

\subsection{Microscopic statistical model approach for cross-section calculation}
Theoretical calculation of reaction cross section requires the construction of an optical-model potential which can efficiently
 describe the absorption (via the imaginary part) as well as scattering (via the  real part). In early days, a phenomenological energy and mass independent mean potential of square-well  Woods-Saxon form was used mostly for astrophysical calculation in local or global form. However, there are many limitations with this prescription and recently there is considerable 
development in microscopic approach.
We have constructed a microscopic  optical-model potential by folding DDM3Y pure nucleon-nucleon interaction with the radial matter density of target to give a direct shape and strength of the nuclear potential. The baryon density  has been 
extracted  from relativistic-mean-field approach.

DDM3Y interaction has  been used  in various works \cite{ddm3y1, ddm3y2} over a wide energy range from few keV to several MeV.
We have applied this optical-model potential in several of our previous works \cite{GG, Chirashree1, Chirashree2, Chirashree3, Chirashree4, Chirashree5, Saumi, Dipti}. It is found that this semi-microscopic potential is well-capable of describing proton capture ($p,\gamma$) reaction rates over  a wide mass region when  real and imaginary potential well depths are being normalized with available  experimental measurements \cite{Chirashree1, Chirashree3, Chirashree5, Saumi, Dipti, n82}.
 The detailed description of the NN interaction and potential formation is given in Dutta \etal \cite{n82}, where we have employed this theory to calculate the $(n,\gamma)$ cross sections for nuclei near $N=82$ neutron  shell closure. In the present work, we calculate the $(n,\gamma)$ reaction cross sections near $N=50$ closed neutron shell.

\begin{figure*}
\includegraphics[scale=1]{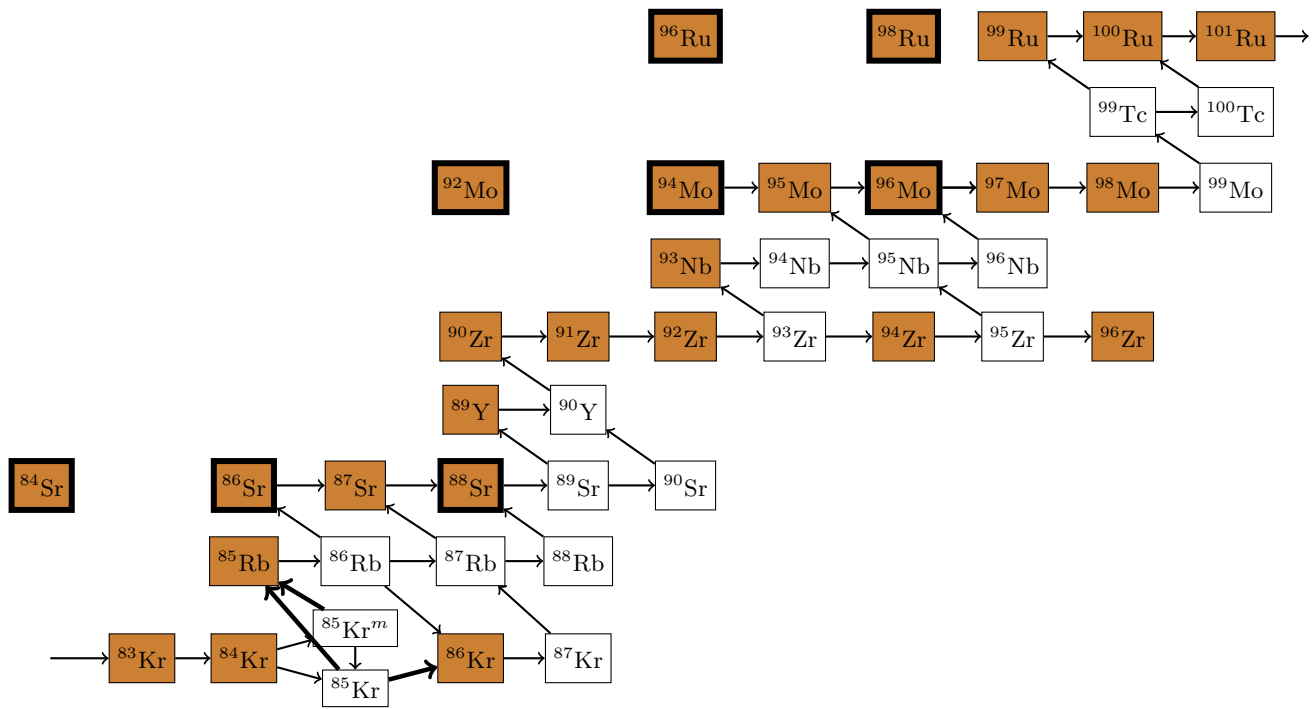}
\label{s-path}
\caption{(Color online) The s-process path near the shell closure at $N=50$. The colored rectangles represent stable and extremely long-lived  isotopes.  The p-nuclei and s-only isotopes are designated by rectangles with thick borders. The strong branching at $^{85}$Kr is shown by thick lines.}
\end{figure*}


The radiative thermal $(n,\gamma)$ cross sections are calculated using statistical model code TALYS1.8 \cite{talys, talys2} applying  DDM3Y nucleon-nucleon interaction folded with radial matter densities of target nuclei obtained from RMF theory. 
Calculation of $(n,\gamma)$ reaction cross sections relevant to s-process in statistical Hauser-Feshbach approach
 depends 
on the choice of various crucial input parameters. The basic ingredient is transmission coefficients for both formation and decay channels of compound nucleus. These coefficients require exact knowledge of nuclear spin, energy, and parity.
One of the basic inputs is nuclear level density. Nowadays, it is believed that  the largest uncertainty in 
reaction cross-section prediction in statistical Hauser-Feshbach approach comes from 
the inappropriate description of nuclear level density \cite{ld_uncert}. Hauser-Feshbach formalism assumes
 a large number of resonances at compound formation energy so that the 
width of individual one can be averaged over the resonances. 
This, in turn, requires a high level density in  the compound nuclear state in the specified 
energy window. We have chosen the level density from Goriely's microscopic calculation \cite{ldmodel} in the combinatorial method that takes into account both rotational enhancement factor as well as vibrational phonon excitations. It is also
 incorporated with appropriate renormalization factors those can efficiently
 reproduce the s- and p-wave neutron resonance spacings. Thus, it has  a further advantage that the data
 can be extrapolated at sufficiently low energies relevant to neutron capture process.

Targets with closed shell, in general, have widely 
spaced nuclear levels. Furthermore, 
 depending upon the level density in the system, different reaction mechanisms  dominate, as different radial parts of the target are probed. 
The reactions near shell closure are characterized by small Q-values. 
Hence, here the cross sections are basically dominated by isolated or narrow resonances. In extreme cases, the direct component can also be important and interference terms may also contribute.  

Another important aspect is given to $\gamma$-ray strength function. 
For radiative capture reactions, the dominant transition, that appears in photon transmission coefficient, is of E1-type. This is  given by, 
\be
T_{E1}=2\pi f_{E1}(E_{\gamma})E_{\gamma}^{3}
\ee 
where, $f_{El}(E_{\gamma})$ is the E1 $\gamma$-ray strength function dependent on $\gamma$-ray energy E$_{\gamma}$ and also on the strength, energy and width of the giant dipole resonances.
It is taken from the microscopic Hartree-Fock-Bogolyubov calculations
 \cite{hartree-fock-bogolyubovgamma} for the present work.

In any reaction mechanism, different channels get activated 
depending upon the energy of the system. These channels have their own partial widths
 and hence the transmission coefficients must be renormalized 
by suitable multiplication factors. This is taken care of by the width 
fluctuation correction. The correction  includes certain correlation factors
  that couple all partial channels altogether. It is thus predominantly
 important near the threshold of new channel openings where channel strengths differ by large factors. In the present study, this correction has been carried out. Pairing energy correction for pre-equilibrium reactions is also included. The number of discrete levels  for decay via a  cascade of $\gamma$ rays is taken to be 30 in number, for both target and residual nuclei.
Full $j,l$ coupling is utilized in HF calculation. Radial densities are taken from RMF model.

\subsection{Energy range of astrophysical importance}
 In the study of neutron induced reactions in astrophysics, one is 
interested in low energy regime, from few keV to several MeV. The $(n,\gamma)$ 
reactions relevant to s-process basically occurs at thermal energies. Unlike charged particles, Gamow peak can not be defined in this case,  as there is no penetration through Coulomb barrier. Hence, for neutron
 induced reactions, location of energy window depends on the contribution from partial waves, (i.e., on the angular momentum quantum number $l$).  Cross sections, in general, can be parameterized as
 a function of energy in terms of dominant partial waves. On this basis,
 a simple approximation is done for energy peak ($E_{0}$) and width ($\Delta$)
 of neutron induced reactions, as follows \cite{energypeakwidth}. 
\begin{equation}
E_{0} \approx 0.172T_{9} \left(l+\frac{1}{2} \right) \text{MeV}
\end{equation}
\begin{equation}
\Delta\approx0.194T_{9} \left(l+\frac{1}{2}\right)^{\frac{1}{2}} \text{MeV}
\end{equation}
Where, $T_{9}$ is the temperature in $GK$. Obviously for s-wave $(l=0)$  neutrons,
energy window is simply given by the  peak and width of Maxwell- Boltzmann distribution. 
For higher partial waves, obviously the peak and width get slightly shifted to higher energies due to penetration
 effect of centrifugal barrier. 
Considering the typical  temperature of  stellar s-process sites, we have calculated $(n,\gamma)$ cross sections
 within the energy range between  1 keV to  1 MeV.

\subsection {Maxwellian-averaged capture cross section (MACS)}
As mentioned earlier that in the astrophysical environment of s-process, cross sections of $(n,\gamma)$ reactions have a spread over a  range of thermal energies.
In the scenario of s-process nucleosynthesis, when a  thermal equilibrium is achieved, the neutron spectrum corresponds to a Maxwell-Boltzmann 
distribution and  Maxwellian-averaged cross sections are obtained by folding the total cross section at particular energy with the thermalized neutron spectra over the wide range of neutron energy. It   is defined as,
\begin{equation}
<\sigma>=\frac{2}{\sqrt\pi} \frac{ \int_{0}^{\infty} \sigma(E_{n}) E_{n} exp(-E_{n}/kT) dE_{n}}{\int_{0}^{\infty} E_{n} exp(-E_{n}/kT) dE_{n}}
\end{equation}
Here $E_{n}$ is the energy in center-of-mass frame.
Thus in principle, MACS should be close to  $\sigma(E_{n})$.
An observation of correlation of MACS values with the size of the nucleus will obviously  show
 local minima at closed neutron shells.

The classical  s-process models usually used the MACS values at a single thermal energy, especially at $KT=30$ keV corresponding to  temperature $3.5\times$10$^{8}$ K. However,  recent stellar models coupled with stellar s-process network codes use  MACSs at different thermal
 energies. Hence, we have calculated MACS values for nuclei having closed neutron shell with $N=50$ for thermal energies ranging from 5 to 100 keV.

\section{Results}
\subsection{Relativistic-Mean-Field Results}
\begin{figure}
\includegraphics[scale=0.065]{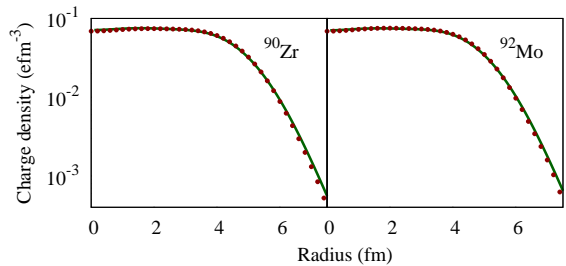}
\caption{(Color online) Charge-density profiles from theoretical relativistic-mean-field model (green solid line)
are compared with 
experimental density distributions obtained from Fourier-Bessel parameter fitting (red points)
of elastic electron scattering distribution \cite{devries}. }
\label{dens_profile}
\end{figure}

\renewcommand{\arraystretch}{1.5}
\begin{table*}[tbh]
\center
\caption{\footnotesize Binding energy (MeV) and charge radius (fm) values extracted from relativistic-mean-field theory are compared with experimental data for nuclei in and around $N=50$ shell closure. Experimental values for binding energy and charge radius are  from Refs. \cite{ame2012} and \cite{angeli}, respectively. 
\label{berad}}
\begin{tabular}{ c rrrr| crrrr | crrrr} 
\hline
Nucleus     & \multicolumn{2}{c} {Binding energy}&   \multicolumn{2}{c|} {Charge radius} &
Nucleus     & \multicolumn{2}{c} {Binding energy}&   \multicolumn{2}{c|} {Charge radius} &
 Nucleus     & \multicolumn{2}{c} {Binding energy}&   \multicolumn{2}{c} {Charge radius} 
\\\hline
  &   Expt.   &      Theory  &        Expt.   &      Theory  &
  &   Expt.   &      Theory  &        Expt.   &      Theory  &     
  & Expt.   &      Theory  &        Expt.   &      Theory  \\
\hline
\hline
$^{82}$Kr&   714.22    &  712.33  &4.192    &        4.142 &
$^{83}$Kr&   721.68     &  720.00   &4.187    &        4.148 &
$^{84}$Kr&   732.22     &  730.74   &4.188    &        4.153 \\
$^{86}$Kr&   749.23     &  747.31   &4.184    &        4.165 &
$^{85}$Rb&   739.24     &  739.20   &4.204    &        4.176 &
$^{84}$Sr&   728.87     &  725.29   &4.239    &        4.189 \\
$^{86}$Sr&   748.88     &  745.98   &4.201    &        4.198 &
$^{87}$Sr&   757.33    &  756.16  &4.225    &        4.203 &
$^{88}$Sr&   768.41     &  766.13   &4.224    &        4.208 \\
$^{89}$Y&   775.46     &  774.24  &4.243    &        4.231 &
$^{90}$Zr&   783.81    &  781.53  &4.269    &        4.254 &
$^{91}$Zr&  791.06    &   787.85  &4.285    &        4.264 \\
$^{92}$Zr&   799.66    &  794.32   &4.306    &        4.274 &
$^{94}$Zr&   814.60    &  806.61   &4.332    &        4.294 &
$^{93}$Nb&   805.75    &  803.15   &4.324    &        4.299 \\
$^{92}$Mo&  796.44     &  794.06   &4.315    &        4.301 &
$^{94}$Mo&  814.23    &  810.77     &4.353    &        4.323 &
$^{95}$Mo&   821.56     &  818.35   &4.363    &        4.334 \\
$^{96}$Ru&  826.46    &  823.83  &4.391    &        4.367 &
$^{98}$Ru&   844.76    &  840.01   &4.423    &        4.389 \\
\hline
\hline
\end{tabular}
\end{table*}

  The density profiles for $^{90}$Zr and $^{92}$Mo with closed neutron shell are shown in Fig. \ref{dens_profile}. In table~\ref {berad},
the theoretical binding energy and charge radius values are listed  with the measured values. Charge radius is the first moment of the nuclear charge distribution. 
Thus, the comparison of root-mean-square (rms) charge radius with experimental 
data can efficiently infer the quality 
of our theoretical RMF model.  It can be easily seen that both our binding energy and charge radius values agree the experimental values very well.  

\subsection{Total capture cross sections}
\begin{figure}
\includegraphics[scale=0.62]{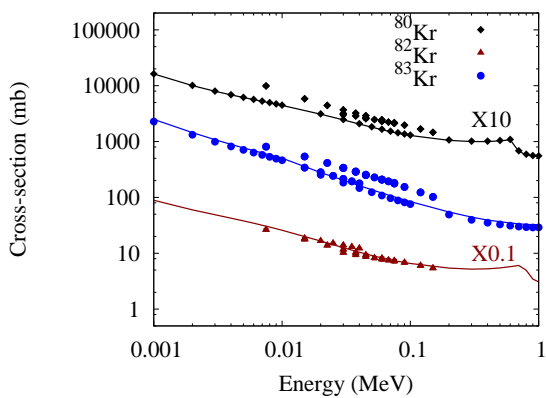}~~

\caption{(Color online) Comparison of results of the present calculation with experimental measurements for $^{80,82,83}$Kr. The  solid lines indicate theoretical results. To avoid overlapping and for the sake of convenience of viewing,  cross-section values of $^{80}$Kr and $^{82}$Kr  have been multiplied by  factors of 10 and 0.1, respectively.
\label{krtotng}}
\end{figure}

\begin{figure}
\includegraphics[scale=0.62]{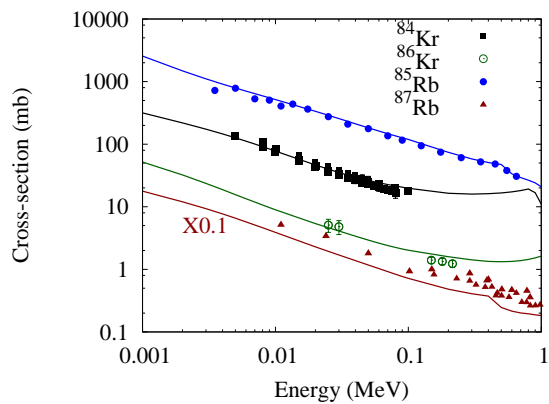}~~

\caption{(Color online) Comparison of results of the present calculation with experimental measurements for $^{84,86}$Kr and $^{85,87}$Rb. The  solid lines indicate theoretical results. To avoid overlapping and for the sake of convenience of viewing, cross-section values of $^{87}$Rb have been multiplied by a factor of 0.1.
\label{krrbtotng}}
\end{figure}

\begin{figure}
\includegraphics[scale=0.62]{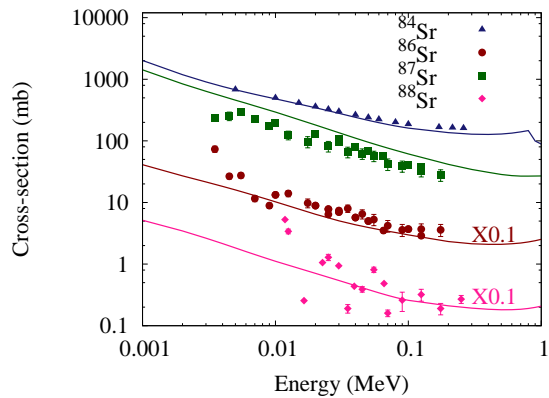}~~
\caption{(Color online) Comparison of results of the present calculation with experimental measurements for $^{84,86-88}$Sr. The  solid lines indicate theoretical results. To avoid overlapping and for the sake of convenience of viewing the cross-section values of $^{86}$Sr and $^{88}$Sr have been multiplied by  factors of 0.1.
\label{srtotng}}
\end{figure}

\begin{figure}
\includegraphics[scale=0.62]{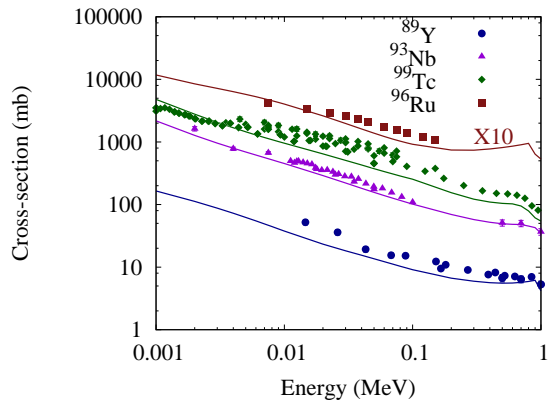}
\caption{(Color online) Comparison of results of the present calculation with experimental measurements for $^{89}$Y and $^{93}$Nb, $^{99}$Tc,  and $^{96}$Ru.
 The  solid lines indicate theoretical results. To avoid overlapping and for the sake of convenience of viewing the cross-section values of $^{96}$Ru have been multiplied by a factor of 10.
\label{ynbrutctotng}}
\end{figure}

\begin{figure}
\includegraphics[scale=0.62]{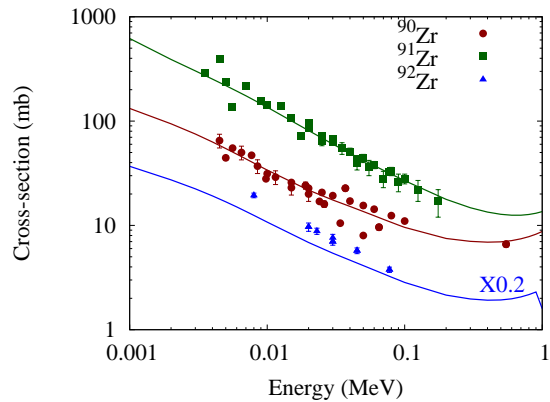}
\caption{(Color online) Comparison of results of the present calculation with experimental measurements for $^{90-92}$Zr.
 The  solid lines indicate theoretical results. To avoid overlapping and for the sake of convenience of viewing,  cross-section values for $^{92}$Zr have been multiplied by a  factor of 0.2.
\label{zr1totng}}
\end{figure}

\begin{figure}
\includegraphics[scale=0.62]{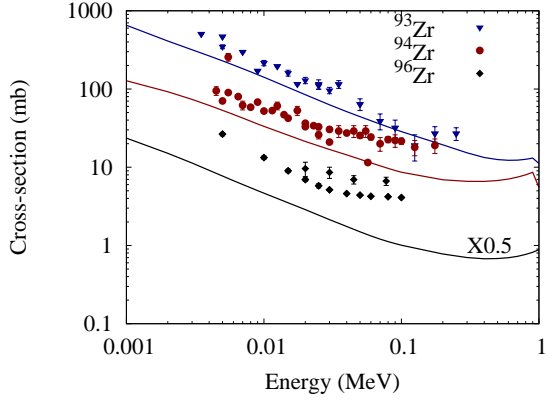}
\caption{(Color online) Comparison of results of the present calculation with experimental measurements for $^{93,94,96}$Zr.
 The  solid lines indicate theoretical results. To avoid overlapping and for the sake of convenience of viewing,  cross-section values for  $^{96}$Zr have been multiplied by a factor of 0.5.
\label{zr2totng}}
\end{figure}

\begin{figure}
\includegraphics[scale=0.62]{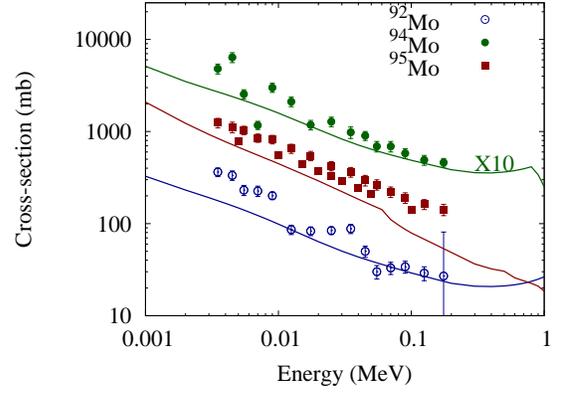}
\caption{(Color online) Comparison of results of the present calculation with experimental measurements for $^{92,94,95}$Mo.
 The  solid lines indicate theoretical results. To avoid overlapping and for the sake of convenience of viewing,  cross-section values for $^{94}$Mo have been multiplied by a factor of 10.
\label{mototng1}}
\end{figure}

\begin{figure}
\includegraphics[scale=0.62]{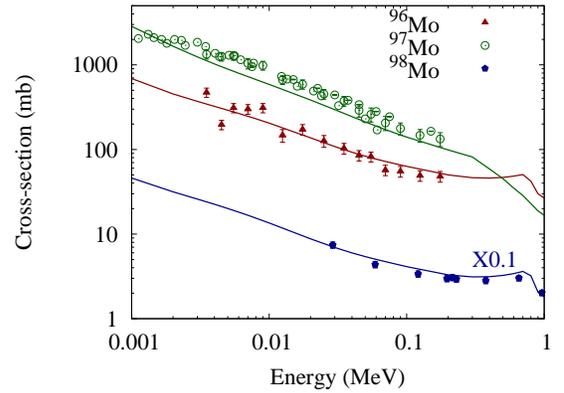}
\caption{(Color online) Comparison of results of the present calculation with experimental measurements for $^{96-98}$Mo.
 The  solid lines indicate theoretical results. To avoid overlapping and for the sake of convenience of viewing,  cross-section values for $^{98}$Mo have been multiplied by a factor of 0.1.
\label{mototng2}}
\end{figure}

Figs.~\ref {krtotng}-\ref{mototng2} show the total $(n,\gamma)$  cross sections on various targets in and around $N=50$ closed neutron shell for 1 keV to 1 MeV  thermal energies. 
The experimental data, in general, are taken   on the basis of most recent measurements and also  where large number of data over a wide energy interval are available. All the experimental data are available at the website of National Nuclear Data Center  \cite{nndcwebref}.
In Figs.~\ref {krtotng} and \ref {krrbtotng}, we plot the experimental and calculated cross sections of $^{80,82,83,84,86}$Kr as well as $^{85,87}$Rb.
Isotopes of krypton are of special interest as they can give valuable information for s-process nucleosynthesis study. Accurate values of cross sections for them are necessary  to eliminate the discrepancies with the isotopic anomalies, i.e., the isotopic  ratios of Kr isotopes. For example, a large spread in the ratio of $^{86}$Kr to $^{82}$Kr with the size of grains found in SiC is a long-standing problem in 
AGB star model of s-process.

The nucleus $^{85}$Kr is a strong branch point.
Production of $^{86}$Kr and  further isotopes of krypton in s-process  is bridged by the neutron
 capture on $^{85}$Kr.  Neutron capture on $^{84}$Kr populates both ground and isomeric states of $^{85}$Kr. The ground  and isomeric states of $^{85}$Kr have  half-lives for $\beta$-decay to $^{85}$Rb of about 10.7 years  and 4.48 hours, respectively. The isomeric state decays to its  ground
state via a $\gamma$-transition by 21\% while  $\beta$-decays to the rubidium isobar by 78\%. At this point, the flow of mass depends on neutron density, i.e., for a sufficiently high neutron density, the neutron channel is open and a competition takes place between $\beta$-decay and neutron capture. The  ground state $\beta$-decay half-life, which stays constant with temperature, becomes long enough so that the s-process path feeds $^{86}$Kr via neutron capture and afterward the path moves towards $^{87}$Rb. It is also possible to the measure the isomeric ratio of $^{85}$Kr if both  total capture cross section and partial capture cross section to the isomer of $^{84}$Kr are known. 

The reaction $^{86}$Kr$(n,\gamma)$$^{87}$Rb is of special interest as $^{86}$Kr  is a closed neutron shell nucleus having comparatively lower cross sections and hence acts as  a waiting point in the reaction  chain. Experimental data for $^{80,82,83,84,86}$Kr are from Refs. \cite{kr80kr82kr83kr84_1, kr84_2, kr80kr82kr83kr84_3, kr82kr83kr84_4,kr86_1, kr86_2}. Mutti \etal \cite{kr80kr82kr83kr84_1} have measured the neutron capture cross sections for several  isotopes of krypton and obtained the MACS values for them in between 5 to 100 keV energy range.
  Raman \etal \cite{kr86_1} identified  several resonances over a large energy range and explicitly took them into account in the determination of total cross sections of $^{86}$Kr. They measured the energy averaged cross sections and concluded that their measurements  suffer from 25\% uncertainty near 30 keV and 30\% uncertainty at 100 keV. 
Beer \etal \cite{kr86_2} measured the average cross sections at mean energies for $^{86}$Kr using activation technique. 
They further  commented on  the contribution from direct p-wave resonance capture for this reaction and suggested the need for more accurate  measurement on this matter.

 The isotope $^{85}$Rb is produced in s-process by the decay of isomeric state of $^{85}$Kr (half-life $\sim$ 4.5 hour) which has 
an overall 80\% probability of $\beta$-decay. Experimental cross sections for $^{85}$Rb is taken from Ref. \cite{rb85_3}.

 The neutron magic isotope $^{87}$Rb also acts as an important bottleneck to the s-process reaction flow. Bao \etal \cite{bao} commented that  large uncertainties exist in the  cross-section values of $^{85,87}$Rb.
Dovbenko \etal \cite{rb87_2} measured the $(n,\gamma)$ cross sections on $^{87}$Rb isotope. They also compared their results with a statistical model calculation using potential well depth of 45 MeV and the compound nuclear  level density parameter
  of 9.4 MeV$^{-1}$. They found a fair agreement below the 1 MeV energy range. However, the spin and parity of 0.847 MeV level for the target was unknown in their calculation which is included in TALYS1.8 database. We have plotted also the data of Dudey \etal \cite{rb87_1} for the same nucleus. They measured the cross sections by activation technique with mono-energetic neutron beam and compared  with theoretical statistical model calculations. They  utilised computer code Abacus \cite{abacus}-Nearrex \cite{nearrex} for their theoretical calculations. Our results are in very good agreement with  the measurement of Dudey \etal while underestimating those of Dovbenko \etal by a factor of $\sim$ 2.
 There are large fluctuations in the results of Beer \etal \cite{rb85_3}
 below 100 keV and we have not  plotted the data here.

Long-lived $^{87}$Rb-$^{87}$Sr chronometric pair can be used for the determination of Galaxy age.
The element  strontium comprises of four isotopes, namely, p-only $^{84}$Sr, a pair of two s-only isotopes $^{86,87}$Sr, and abundant magic neutron isotope $^{88}$Sr. 

Fig. \ref{srtotng} shows the $(n,\gamma)$ cross sections on Sr targets. Data for the p-nucleus $^{84}$Sr is taken
 from Dillmann \etal \cite{sr84}. They performed  measurement at $kT=25$ keV via activation technique and reported an overall 
5.7\% error in their measurement with the major
 uncertainty coming from $\gamma$-ray intensity. Further, they extrapolated the cross sections at various thermal energies from 5 to 260 keV by normalising the data with JEFF-3.0 \cite{jeff3.0}, NON-SMOKER \cite{nonsmoker1,nonsmoker2}, JENDL-3.3 \cite{jendl3.3}, and ENDF/-VI.8 \cite{endfb6.8} to derive MACS values.
 The agreement of their data with our theory is extremely good, 
especially for energies below 50 keV.

Cross sections on $^{86,87}$Sr are important for s-process studies
 as they  experience full s-process flow in the network. They can as well serve the purpose of cosmo-chronometry. They are produced together  due to the  branchings  at $^{85}$Kr and $^{86}$Rb. Hence, the ratio of  $^{87}$Sr to $^{86}$Sr should remain close to the inverse of their respective cross-section ratios.
Bauer \etal \cite{sr86sr87} reported that cross sections for these two particular reactions have to be known to an accuracy of 5\% or better. They measured $(n,\gamma)$ cross sections
 on $^{86}$Sr and $^{87}$Sr targets for energies ranging from 100 eV to 1 MeV using TOF facility.
They found strong resonance structures for both $^{86}$Sr and $^{87}$Sr below 10 keV. Suitable corrections were made and no averaging  over energy bins was done by them. Their data decreases according to $1/v$ law above 20 keV. Our results fairly reproduce their data. Our calculation overestimates those of Hicks \etal \cite{sr87hicks} for  $^{87}$Sr. However, there remains certain ambiguity with their data for $^{86}$Sr at extremely low energies below 7 keV. Macklin and Gibbons \cite{sr86sr87_4} presented two data for both  $^{86}$Sr and $^{87}$Sr below 100 keV neutron energy. The experimental data of abundant neutron magic $^{88}$Sr are taken from Refs. \cite{sr88_1,sr88macs30}. This nucleus is mostly from s-process origin. The data are highly scattered and suffer from large uncertainties, as can be seen from Fig. \ref{srtotng}.

The data for $^{89}$Y are taken from Refs. \cite{y89_1nb93_2, rb85_1rb87_1y89_2}. 
These are plotted with our calculations in Fig. \ref{ynbrutctotng}. This nucleus with the odd mass number and the magic neutron number is of particular interest as it has  very low neutron capture cross sections and thus acts as a bottleneck in s-process reaction flow. Stupegia \etal \cite{rb85_1rb87_1y89_2} compared their data with statistical model calculations and adjusted the ratio of radiation width to the observed level spacing which was taken as a significant 
 parameter in their calculation to obtain  a reasonable fitting. It can be seen from Fig. \ref{ynbrutctotng}
 that our results agree with   both measurements quite well.

 In Figs.~\ref {zr1totng} and \ref{zr2totng}, we plot the cross sections of $^{90-94, 96}$Zr 
with experimental data. Zirconium does not have any s-only or p-only isotopes. The isotopes $^{90-94}$Zr are mostly of s-process origin. The isotope $^{96}$Zr is believed to be from both s- and r-process  because of small half life of $^{95}$Zr (64.032 days).
 For $^{90,91,92}$Zr the data are from Refs. \cite{zr90,zr91zr92,zr91_2,zr91_3,zr92_2}.
Tagliente \etal \cite{zr92_2}  measured Maxwellian-averaged cross sections for $^{91,92}$Zr with improved n\_TOF  method.
The experimental determination of small  $(n,\gamma)$ cross sections for $^{93}$Zr, which is an important LLFP, 
 is somewhat difficult due to its radioactive nature with a sufficiently large $\beta$-decay half-life of about 1.53$\times10^{6}$ years. The sample of this isotope contains very poor enrichment and hence high-resolution TOF measurement is required in this regard.
 R. L. Macklin \cite{zr93_1}  derived the average neutron capture cross sections 
for this nucleus using TOF  technique from 3 to 300 keV. He identified 138 resonance peaks and further calculated the Maxwellian-averaged cross sections for a range of thermal energies from 5  to 100 keV. He compared their measurement with that of ENDF/B-V evaluation \cite{zr93endfbcomp} and that of Ref. \cite{iijima}. While the data of Iijima  \etal \cite{iijima} lie well above his measured values, the data of ENDF/B evaluation agrees well at higher energies but lie slightly below the experimental results for energy below 60 keV. Our calculations
 are in good agreement  with those of  Macklin  \cite{zr93_1}, especially at higher energies.
 Tagliente \etal \cite {zr93_2} have also measured the Maxwellian-averaged cross sections for this reaction with n\_TOF collaboration. The final MACS values were obtained  by renormalizing  their values with JENDL-4.0 library data \cite{JENDL} to eliminate the discrepancy 
for small capture kernels.

Cross sections for $^{94,96}$Zr are required the in  the analysis of s-process branching at $^{95}$Zr. Similar to the branching at $^{85}$Kr, this branching is also independent of stellar temperature and hence, the isotopic ratio of $^{94}$Zr to $^{96}$Zr can predict the stellar neutron density condition. The two isotopes $^{94,96}$Zr suffer from overabundance problem. The s-process also contributes slightly to the production of $^{96}$Zr, which, in general, believed to be r-only isotope \cite{kappeler2}, due to the branching at $^{95}$Zr at high neutron densities. Hence, accurate determination  of capture cross sections is needed 
to overcome this problem.  The experimental data for $^{94}$Zr are from Refs. \cite{zr94_2, zr94_3, zr9496exp_1} and for $^{96}$Zr are from Refs. \cite{zr9496exp_1, zr96exp_2}.
According to Bao and K{\"a}ppeler \cite{bao1987}, existing experimental neutron capture cross sections
 for $^{96}$Zr exhibit large discrepancies up to a factor of 2. The isotope $^{96}$Zr, once produced, is only slightly destroyed by neutron capture because of its low neutron capture cross sections. Thus, the accurate cross section is required to indicate the efficiency of neutron source during s-process nucleosynthesis.  Our theory underpredicts the measurements by  a factor  on an average $\sim$ 3.


The isotope $^{93}$Nb is produced via the decay of  long-lived radioisotope $^{93}$Zr. It  is then fully converted to $^{94}$Nb during s-process nucleosynthesis. Several measurements exist for  $(n,\gamma)$ cross sections on $^{93}$Nb. We have taken those 
from Refs. \cite{nb93_1, y89_1nb93_2, nb93_3}. Xia \etal \cite{nb93_3} have measured the same using a set-up of 
Moxon-Rae detectors in the neutron energy range from 1 to 60 keV. It can be easily seen from Fig. \ref{ynbrutctotng} that our results for this isotope fairly agree with all of them.

\setlength{\tabcolsep}{12pt}
\renewcommand{\arraystretch}{2.0}
\begin{table*}[htb]
\caption{Maxwellian-averaged cross sections at $kT=30$ keV for nuclei near the 
$N=50$ shell closure. Experimental values are from Ref. \cite{kadonis}. For unstable and radioactive nuclei, experimental data are not available. See text for other 
experimental values. The nuclei with $N=50$ are in bold font.
\label{macs30kev}}
\begin{tabular}{crrrcrrr}\hline
 &\multicolumn{3}{c}{MACS (mb)}&
 &\multicolumn{3}{c}{MACS (mb)}\\\cline{2-4}\cline{6-8}
Nucleus&Pres. & Exp. & MOST & Nucleus&Pres. & Exp. & MOST\\\hline
$_{36}^{80}$Kr&     245  &    267$\pm$14   &  223&
$_{36}^{82}$Kr&     130  &    90$\pm$6    &  112\\
$_{36}^{83}$Kr&     208  &    243$\pm$15    &  170&
$_{36}^{84}$Kr&     38.3  &    38$\pm$4    &  58.9\\
$_{36}^{85}$Kr&     40.1  &       &  51.3 &
{\bf $_{36}^{86}$Kr}&   4.35 &    3.4$\pm$0.3    & 3.71 \\
$_{37}^{85}$Rb&     259  &    234$\pm$7    & 197 &
$_{37}^{86}$Rb&     206 &       &  226\\
{\bf $_{37}^{87}$Rb}&     17.6  &    15.7$\pm$0.8    &  18.8&
$_{38}^{84}$Sr&     283  &    300$\pm$17    & 244 \\
$_{38}^{86}$Sr&     54.5  &    64$\pm$3    & 51.0 &
$_{38}^{87}$Sr&     139  &    92$\pm$0.4    &72.2  \\
{\bf $_{38}^{88}$Sr}&     5.55  &    6.13$\pm$0.11 &5.02  &
$_{38}^{89}$Sr&     17.6  &        &  22.0\\
$_{38}^{90}$Sr&     7.02  &        &  &
{\bf $_{39}^{89}$Y}&     19.0  &    19$\pm$0.6    &16.6  \\
{\bf$_{40}^{90}$Zr}&     18.4  &    19.3$\pm$0.9   &13.7  &
$_{40}^{91}$Zr&     62.3 &    62.0$\pm$3.4  &53.7  \\
$_{40}^{92}$Zr&     28.1  &    30.1$\pm$1.7    &25.5 &
$_{40}^{93}$Zr&     65.3  &    95$\pm$10    &67.8 \\
$_{40}^{94}$Zr&     17.4  &    26$\pm$1    &13.3  &
$_{40}^{95}$Zr&     31.9  &         & 29.1 \\
$_{40}^{96}$Zr&     4.46  &    10.7$\pm$0.5    & 10.7 &
$_{41}^{93}$Nb&     224        &    266$\pm$5    &  241\\
$_{41}^{94}$Nb&     501  &      &  377 &
$_{41}^{95}$Nb&     94.6  &        &  112\\
{\bf $_{42}^{92}$Mo}& 53.2  &    70$\pm$10 &  45.5    &
$_{42}^{94}$Mo&     87.2  &    102$\pm$20    & 84.2 \\
$_{42}^{95}$Mo&     212  &    292$\pm$12    &237  &
$_{42}^{96}$Mo&     113  &    112$\pm$8    &112  \\
$_{42}^{97}$Mo&     299  &    339$\pm$14    &276 &
$_{42}^{98}$Mo&     73.8  &    99$\pm$7   &58.2  \\
$_{42}^{99}$Mo&     262  &  &  337&
$_{43}^{99}$Tc&      513    &    933$\pm$47      & 423  \\
$_{44}^{96}$Ru&      197    &    207$\pm$8      & 335  &
$_{44}^{98}$Ru&      238   &         & 355  \\
\hline
\end{tabular}
\end{table*}

\setlength{\tabcolsep}{1.5pt}
\renewcommand{\arraystretch}{1.2}
\begin{table}[hbt]
\caption{MACS values from a few of other selected works at various energies are listed with our results.
\label{macsothers}}
\begin{tabular}{cccc rr}
\\
\hline
Nucleus &  Energy (MeV) &
 \multicolumn{3}{c}{Cross section (mb)} \\ \hline
&&  Present  & \multicolumn{2}{c}{Ref.}  \\ \cline{3-5}
$^{84}$Kr   &0.003&"& 36.3$\pm$2&\cite{mughabghabmacsrecom}\\
$^{86}$Kr &0.020&Table \ref{macs}&6.6 &\cite{kr86macswalter}\\
& 0.025 & "&3.49$\pm$0.4 & \cite{kr86macsbeer}& \\
   &"&"& 5.1&\cite{kr86_1}\\
&"&"&3.58$\pm$0.3&\cite{kr86macskappeler}\\
   &0.030&Table~\ref {macs30kev}, \ref{macs}& 4.1$\pm$0.3&\cite{mughabghabmacsrecom}\\
&"&"&5.7 &"\\
&0.040&Table \ref{macs}&5.1 &"\\
&0.050&"&4.25&"\\
&0.052&3.12&3.30$\pm$0.5&"\\
$^{85}$Rb&0.025&290&234&\cite{heilmacs}\\
$^{87}$Rb&0.012&33.8&26.3$\pm$2.6&\cite{jaagrb87}\\
&0.020&Table~\ref{macs}&19.5$\pm$2.0&"\\
&0.030&Table~\ref {macs30kev}, \ref{macs}&15.5$\pm$1.5&"\\
&0.040&Table~\ref {macs30kev}&13.4$\pm$1.3&"\\
&0.050&"&12.4$\pm$1.2&"\\
$^{84}$Sr&0.030&"&370 $\pm$17&\cite{mughabghabmacsrecom}\\
$^{86}$Sr&0.030&"&57 &\cite{sr86macs30}\\
&0.030&"&56.9 &\cite{kadonis}\\
$^{87}$Sr&0.030&"&108$\pm$20&\cite{sr86sr87_4}\\
$^{88}$Sr&0.025&Table~ \ref{macs}&6.72 $\pm$0.18&\cite{sr88y89Kappelermacs}\\
&0.030&Table~\ref {macs30kev}, \ref{macs}&5.6 &\cite{sr88macs30}\\
&"&"&5.59 &\cite{kadonis}\\
$^{90}$Zr&"&"&11$\pm$3&\cite{macklinzr}\\
$^{91}$Zr&"&Table~\ref {macs30kev}&59 $\pm$10&\cite{macklinzr}\\
$^{92}$Zr&"&"&34 $\pm$6&"\\
$^{94}$Zr&"&"&21 $\pm$4&"\\
$^{96}$Zr&"&"&41 $\pm$12&"\\
\hline
\end{tabular}
\end{table}

\setlength{\tabcolsep}{7pt}
\renewcommand{\arraystretch}{1.5}
\begin{table*}
\caption{Maxwellian-averaged neutron capture cross sections of
$^{86}$Kr, $^{87}$Rb, $^{88}$Sr, $^{89}$Y, $^{90}$Zr, and $^{92}$Mo.
The experimental values  are from Ref. \cite{kadonis}. 
\label{macs}}
\begin{tabular}{crr|rr|rr|rr|rr|rr}\hline
Energy (MeV) & \multicolumn{12}{c}{MACS (mb)}\\
\hline
&\multicolumn{2}{c}{$^{86}$Kr}&\multicolumn{2}{c}{$^{87}$Rb}
&\multicolumn{2}{c}{$^{88}$Sr}&\multicolumn{2}{c}{$^{89}$Y}
&\multicolumn{2}{c}{$^{90}$Zr}&\multicolumn{2}{c}{$^{92}$Mo}\\
\hline
& Pres. & Expt.& Pres. & Expt.& Pres. & Expt.
& Pres. & Expt.& Pres. & Expt. &Pres.& Expt.\\
\hline
0.005& 15.6 &2.1  &63.6  &  111.2&18.0&10.88&64.7&68&52.7&44.4&144 &277\\
0.010& 9.30& 3.2  &38.6 &    63.2&11.3&11.86&40.3&40&34.2&31.3&96.4&158\\
0.015& 6.93& 3.6  &28.7 &    48.1&8.63&9.88&30.8&30&26.9&25.8&76.6&115\\
0.020& 5.67& 3.7  &23.4 &    40.3&7.16&8.21&25.6&25&22.8&22.7&65.5 &93\\
0.025& 4.89& 3.5  &19.9 &    35.7&6.22&7.02&22.3&21&20.2&20.7&58.4 &79\\
0.030& 4.35& 3.4  &17.6 &    32.4&5.55&6.13&19.0&19&18.4&19.3&53.2 &70\\
0.040& 3.64& 2.9  &14.3 &    27.7&4.67&5.04&16.7&16&15.8&17.1&46.2 &59\\
0.050& 3.19& 2.5  &12.2 &    23.6&4.12&4.35&14.7&14&14.1&15.5&41.6 &53\\
0.060& 2.88& 2.2  &10.8 &    22.0&3.73&3.95&13.2&13&13.0&14.3&38.3&49\\
0.080& 2.49& 1.8  &8.87 &    18.4&3.25&3.25&11.3&11.3&11.4&11.4&34.0&45\\
0.100& 2.26& 1.5  &7.67 &    15.2&2.96&3.36&10.5&10&10.5&11.0&31.4 &43\\
\hline
\end{tabular}
\end{table*}

There is a series of isotopes of molybdenum those efficiently take part in heavy element  nucleosynthesis. The two isotopes $^{92,94}$Mo are p-only with $^{92}$Mo  being  neutron magic. While the isotope $^{96}$Mo is s-only, $^{95,97,98,99}$Mo are produced partly in s-process and partly in r-process, and $^{100}$Mo is r-only. Results  for  $(n,\gamma)$ cross sections for $^{92,94-98}$Mo are shown in Figs. \ref {mototng1} and \ref{mototng2}. The data for $^{92,94-96,97}$Mo are taken from 
Musgrove \etal \cite{mo92mo94mo95mo96}. They measured the radiative capture cross sections averaged over energy bins
 on several molybdenum isotopes  from 3 to 90 keV with high-resolution TOF technique and extracted s- and p-wave 
resonance parameters. As can be seen from the figure that their values are in good agreement with our calculations. We have also plotted the data of Kapchigashev \etal \cite{mo97_1} for $^{97}$Mo. The data for $^{98}$Mo are taken from Chunhao \etal \cite{mo98}. The results are in good agreement for this isotope.

We have also plotted the $(n,\gamma)$  cross  sections for $^{99}$Tc in Fig. \ref{ynbrutctotng}. The experimental values are from Refs. \cite{tc99_1,tc99_2,tc99_3,tc99_4}. The most recent measurement has been done by Matsumoto \etal \cite{tc99_2} in TOF technique. They have measured for incident neutron energies from 8 to 90 keV and at 190, 330, and 540 keV and obtained the results with an error of about 5\%.
Rapp \etal \cite{ru} measured the  neutron-induced capture cross sections on Ru isotopes including 
 p-only nucleus $^{96}$Ru using activation technique at $kT=25$ keV. The extrapolation is done at higher 
and lower energies by normalizing the data of Bao \etal \cite{bao}. They are in fair  agreement with the present calculations as can be seen from Fig.~\ref{ynbrutctotng}.

\subsection{Maxwellian-Averaged Cross sections (MACS)}
In Table~\ref{macs30kev},  MACS values calculated in the present approach  are presented along with experimental values,  and theoretical
 values from MOST calculations, for nuclei relevant to s-process in and around $N=50$ closed shell including p-only and s-only isotopes.  The experimental values are available at the KADoNis database \cite{kadonis}, which is an updated version of the compilation by Bao \etal \cite{bao}.
Some other theoretical calculations on MACS values are also available. However, 
most of them suffer from large uncertainties and hence do not agree well with the measured ones. The reason for variations in the existing theoretical MACS data is due to different choices of nuclear parameters which enter into the calculations.   MOST calculations have been  performed 
under microscopic Hauser-Feshbach  approach taking  JLMB nucleon-nucleon interaction potential \cite{JLMB} into account. In
 Table \ref{macsothers}, we have also listed MACS values from some selected other theoretical and experimental works at various energies and compared them with our present results.
 Below we will discuss certain important aspects of MACS values relevant to this work. 

Recommended values for $^{84}$Kr and $^{86}$Kr by S. F. Mughabghab  \cite{mughabghabmacsrecom} lie extremely close to
 our MACS values  at 30 keV. 

There exists considerable uncertainty in the existing MACS values of $^{85}$Rb. We have also presented MACS at 25 keV for this 
isotope in Table~\ref{macsothers}.
 No experimental MACS  is available for unstable $^{86}$Rb. Our calculation yields a value of  206 mb at 30 keV. The   theoretical value, as recommended by Bao \etal \cite{bao}, is 202$\pm$163 mb. The magnitude of this MACS value suggests the possibility of a weak branching at this point.
 It is mentioned previously that the long-lived neutron magic $^{87}$Rb (half-life=4.81$\times$10$^{10}$ years) acts as bottleneck due to its low cross-section value (Table \ref{macs30kev}). 

For $^{84}$Sr, S. F.  Mughabghab  \cite {mughabghabmacsrecom} recommended a value as 370$\pm$17 mb at 30 keV.
It can be seen from Table \ref{macsothers} that our result for $^{86}$Sr agrees the theoretical calculation of Harris \cite{sr86macs30}
   as well as   the theoretical
 calculation of  MOST2002  very well. The reported MACS value by Macklin and Gibbons \cite{sr86sr87_4} for $^{87}$Sr at 30 keV,
 as given in Table \ref{macsothers}, is  close to our result.  Excellent agreements have  been obtained with the measurement of Boldeman \etal \cite{sr88macs30} as well as theoretical MOST2002 calculation at 30 keV for $^{88}$Sr, given in Table~\ref{macsothers}.
 Minor branchings can occur at the two isotopes $^{89,90}$Sr with $\beta$-decay half-lives of 50.53 days and 28.90 years, respectively. Hence, we have given the MACS values for them at an energy of 30 keV in Table \ref{macs30kev}. Bao \etal \cite{bao} have recommended a theoretical MACS value of 19$\pm$14 mb at 30 keV for $^{89}$Sr.

Experimental MACS values for $^{90,91,92}$Zr, as recommended by Bao \etal \cite{bao},  are very well reproduced. Further, the other existing measurements for $^{91,92}$Zr \cite{kadonis} are also found to correspond closely with our obtained results  within the quoted uncertainties.
While that of $^{93}$Zr agrees well 
with MOST prediction, it comes out to be  lower than the measured one. Similarly, 
cross sections for $^{94,96}$Zr  are underproduced by the current semi-microscopic approach.
 It is interesting to note that theoretical MOST
 calculations also found much smaller values for the two isotopes. 
 Macklin  and Gibbons \cite{macklinzr} measured the MACS values for several isotopes of Zr. They are listed in Table \ref{macsothers}.
  Obviously,  the one for $^{96}$Zr corresponds very poorly with our result.

The existing experimental  data in the literature on $^{93}$Nb vary significantly in magnitudes \cite{kadonis}.
The radioactive nucleus $^{94}$Nb  has a long
 $\beta$-decay half-life (2.03$\times$10$^{4}$ years). No experimental signature is available for neutron cross section on it. Our theory yields a MACS value of 501 mb at 30 keV. Thus, the $(n,\gamma)$ cross section at this energy  is quite large that prevents its decay to $^{94}$Mo which is produced entirely in p-process. A possible branching may  occur at $^{95}$Nb (half-life=34.99 days) if there exists a high neutron flux at s-process temperatures and hence we have provided the MACS  for this isotope.

Our values for molybdenum isotopes are in satisfactory agreement with experimental values. 
Minor branchings can also be possible at $^{99}$Mo (half-life=65.976 hours) and long-lived $^{99}$Tc (half-life=2.111$\times$10$^{5}$ years) and hence we have presented MACS values for them in TABLE \ref{macs30kev}. Bao \etal \cite{bao} have recommended a theoretical value of 240$\pm$40 mb for $^{99}$Mo at 30 keV, which corresponds to our calculation within the quoted errors.


The long-lived LLFP $^{99}$Tc has a large $(n,\gamma)$ cross section. Other than the recommended value of Bao \etal \cite{bao} which is from the measurement of Gunsing \etal \cite{tcmacs1},  other two TOF measurements on MACS are available which presented the values 782$\pm$ 50 mb \cite{tcmacs2} and 779$\pm$40 mb \cite{tc99_4} at 30 keV.
Our MACS at 30 keV for p-nucleus $^{96}$Ru agrees recent measurement of
 Rapp \etal \cite{ru}  far better in respect of all existing theoretical and old measurement values. 

For $^{85}$Kr, $^{86}$Rb, $^{89,90}$Sr, $^{95}$Zr,  $^{94,95}$Nb, $^{99}$Mo, and $^{98}$Ru.
 no experimental MACS value is available. However,  several theoretical estimates, largely 
differing from each other, can also be found in literature.  

The modern stellar codes on s-process
 nucleosynthesis demand  MACS values at different energies rather than a single energy of
 30 keV. Hence in table~\ref{macs}, we present the MACS values at various thermal energies
 for nuclei involved in astrophysical s- and p-processes containing the magic number ($N=50$) of
  neutrons.

 Some experimental measurements are also available at  energies other than 30 keV. 
As discussed previously, Mutti \etal \cite{kr80kr82kr83kr84_1} obtained the MACSs from 5 to 100 keV for several isotopes of krypton. They found a systematic discrepancy with the data of Refs. \cite{kr86macskappeler,beerkr} measured with activation techniques and of Refs. \cite {kr86_1,walterkr86} measured with TOF techniques for $^{86}$Kr. 
Tagliente \etal \cite{zr96_2zr90macstagliente}  estimated the MACS values for $^{90}$Zr from 5 to 100 keV
  by folding the total capture cross sections measured via  n\_TOF method with  the
 neutron spectra at thermal equilibrium.  Their results agree  our calculations  very well.
  Heil \etal \cite{heilmacs}  evaluated MACS
 values for $^{87}$Rb from 5 to 100 keV by normalizing the  neutron spectrum at 25 keV with 
 JENDL-3.3 \cite{jendl3.3}, JEFF-3.1 \cite{jeff3.1}, and ENDF/B-VI.8 \cite{endfb6.8} library data. Our theory reproduces their work very well
 in between 20 to 100 keV region.


\section{summary}
We have constructed a semi-microscopic optical-model potential to calculate  thermal 
$(n,\gamma)$ reaction cross sections, in which, the target after absorbing the neutron,
 emits one or more $\gamma$ rays, for energies ranging from 
1 keV to 1 MeV and  for nuclei in and around $N=50$ closed neutron shell. The nuclei
 involved are of astrophysical interests
 taking part in s-process and p-process nucleosynthesis.  Standard DDM3Y NN interaction,
 folded with target matter densities, obtained from RMF theory is used. The calculations 
are done in fully statistical Hauser-Feshbach approach
 with  the standard reaction code TALYS1.8. We have compared the results with the available
 experimental data.  Our theory reproduces most of the measurements with reasonable 
agreement. The experimental data in some cases are extremely old and hence suffer
 from underestimation of proper background and error correction. Also some significant 
inputs in statistical calculations suffer from the lack of measured data.
For example, experimental signature of  giant dipole resonance (GDR) for
 $^{86}$Kr is not available yet and hence enters into the statistical model 
calculations from systematics. Thus, unavailability of experimentally measured 
parameters is a possible reason for the discrepancies
 between experiments and statistical model predictions.  Further,
The Maxwellian-averaged cross sections are calculated  and presented
 with experimental measurements. We have also listed the statistical MOST2005
 predictions of MACS values.
 

\section{Acknowledgment}
The authors acknowledge the financial support from University Grants
 Commission (JRF and DRS), Department of Science and Technology, and Alexander Von Humboldt Foundation.

\end{document}